\documentstyle[11pt,moriond,epsfig]{article}

\def\Journal#1#2#3#4{{#1} {\bf #2}, #3 (#4)}

\def\PLB{{\em Phys. Lett.}  B}
\def\PRL{\em Phys. Rev. Lett.}
\def\PRD{{\em Phys. Rev.} D}
\def\PRC{{\em Phys. Rev.} C}

\def\be{\begin{equation}}
\def\ee{\end{equation}}
\def\bea{\begin{eqnarray}}
\def\eea{\end{eqnarray}}

\begin{document}
\vspace*{4cm}
\title{CAN LSND AND SUPERKAMIOKANDE BE EXPLAINED BY RADIATIVE DECAYS OF $\nu_{\mu}$'S?}

\author{ F. Vannucci }

\address{LPNHE, Universit$\acute{e}$ Paris 7,\\
Paris, France}

\maketitle
\abstracts
{The radiative decay of $\nu_{\mu}$'s in matter with a scheme of 
mass-degenerate neutrinos could be the common origin of the appearance of 
$\bar{\nu_{e}}$'s at LSND and the disappearance of $\nu_{\mu}$'s at 
SuperKamiokande. With the decay probability fixed by the LSND signal, 
the deficit of atmospheric neutrinos can be satisfactorily reproduced.}

\section{Introduction}

There are several puzzles in neutrino physics:
\begin{itemize}
\item the appearance of $\bar{\nu_{e}}$'s in the LSND experiment, not confirmed
by the very similar experiment Karmen \cite{LSND_Karmen}.

\item the disappearance of atmospheric $\nu_{\mu}$'s at SuperKamiokande over 
distances of the order of the earth's diameter \cite{SuperK}.
\end{itemize}
These two findings have been interpreted as evidence of neutrino oscillations. 
Together with the solar deficit also interpreted as a sign of oscillations, 
it is difficult to build a coherent scenario with the only three neutrinos 
which are known to exist.

Another possibility is considered here, namely the radiative decays of 
$\nu_{\mu}$'s in the context of mass-degenerate neutrinos, for example 
neutrinos having masses of a few eV for cosmological purposes and related 
by a $\delta m^2$ fixed by the solar deficit. This could explain both the 
LSND and SuperKamiokande signals.

 Decays of neutrinos have been advocated \cite{Nudecay} and rejected 
\cite{Nudec_rej} as a solution for the atmospheric deficit. We consider 
here the radiative mode which is hugely amplified by matter effects 
\cite{matter_effect1,matter_effect2,matter_effect3}. This process differs 
from the simple case of decays in vacuum considered up to now in two aspects: 
antineutrinos may not be affected (the refraction index is different for 
neutrinos and antineutrinos), and the decay probability varies rapidly with 
the density of the traversed medium.

\section{Interpretation of the LSND signal}

The radiative decay of neutrinos consists of the process:\\
  $$\nu_2 \rightarrow \nu_1 + \gamma$$

where $\nu_2$ and $\nu_1$ are mass eigenstates, $\nu_2$ being the heaviest 
one. In a simple scheme, $\nu_2$  is predominantly $\nu_{\mu}$ and $\nu_1$ 
predominantly $\nu_e$. As a consequence of the helicity flip in the transition,
the final neutrino is right-handed. If neutrinos are Dirac particles, the 
emerging neutrino is sterile. If, on the other hand, neutrinos are Majorana 
particles, the right-handed final neutrino is active and the process can be 
written:

  $$\nu_{\mu} \rightarrow \bar{\nu_{e}} + \gamma$$

 This is the decay mode which will be assumed for the present argument. 
Similar considerations of stimulated conversion between mass-degenerate 
neutrinos have been discussed \cite{stim_conv}.

Radiative decays of $\nu_{\mu}$'s have been searched for experimentally 
\cite{rad_dec_exp}. The result is $\tau$/m $\geq$ 15.4 s/eV, where m is 
the mass of the decaying neutrino. This result seems to exclude the 
considerations which are developed below. However this limit only applies 
to neutrinos with very different masses, when the emitted photon takes 
half of the incident neutrino energy. With mass-degenerate neutrinos, 
the limit does not apply, and the $\bar{\nu_{e}}$ takes up most of the 
incident energy. This process could therefore be at the origin of the LSND 
signal.
 
The LSND beam is composed of $\nu_{\mu}$, $\bar\nu_{\mu}$ and $\nu_{e}$ at 
equal level, but contains almost no $\bar\nu_{e}$. A signal of $\bar\nu_{e}$ 
is claimed and the favoured interpretation is the oscillation of 
$\bar\nu_{\mu}$ into $\bar\nu_{e}$ . The decay discussed above would be 
equally satisfactory. In fact, it would explain why the Karmen experiment 
does not see a signal. With Karmen, the beam is better time-defined, the 
$\nu_{\mu}$'s and $\bar\nu_{\mu}$'s are well separated, and the oscillation 
is specifically searched from the $\bar\nu_{\mu}$ component.

If this is the correct interpretation of the LSND signal, it gives a decay 
probability of $3 \ 10^{-3}$ for 30 MeV neutrinos, over a decay path of about 
30 m (distance between the beam stop and the centre of the detector). With 
these parameters the lifetime is:
$\tau$/m $\simeq 10^{-12}$ s/eV.

Such a short lifetime is not a priori excluded by laboratory limits, which 
only apply to non-degenerated neutrino masses.

\section{Consequences for atmospheric neutrinos}

Let us now consider a 1 GeV $\nu_{\mu}$ travelling along a flight path of 
13000 km (diameter of the earth). This is the typical situation encountered 
with atmospheric neutrinos. The lifetime inferred from LSND gives a 
$\gamma$c$\tau$ of $3 \ 10^5$ m. This is more than an order of magnitude too 
small to give a decay probability corresponding to the level of disappearance 
seen by the SuperKamiokande experiment for upward going neutrinos.

However, the case to be considered is more complex, as the neutrinos are 
travelling through matter. It has been shown that radiative decays of 
neutrinos are hugely amplified in dense media. The lifetime $\tau_m$ in 
matter is related to the lifetime in vacuum $\tau_0$ by the expression:

$$\Large \frac{\tau_0}{\tau_m} \normalsize = 8.6 \ 10^{23}F(v) \Large (\frac{N_e}{10^{24}cm^{-3}})^2 (\frac{1eV}{m})^4$$

where $N_e$ is the electron density of the medium. This formula applies 
for neutrinos with a mass hierarchy. For mass-degenerate neutrinos, it becomes:

$$\Large \frac{\tau_0}{\tau_m} \normalsize = 8.6 \ 10^{23}F(v) \Large (\frac{N_e}{10^{24}cm^{-3}})^2 (\frac{1eV}{m})^4(\frac{m^2}{\delta m^2})^2$$

The value of F(v) has not been completely elucidated. For relativistic 
neutrinos the term F(v) tends to 4 m/E according to some authors \cite{matter_effect2} whilst it is 
about 1 according to others \cite{matter_effect3}. The issue needs further calculations, 
and we have adopted the naive approach, with an amplification proportional 
to the square of $N_e$, and inversely proportional to the neutrino energy. 
Taking into account these factors, let us reconsider the cases of LSND and 
SuperKamiokande.

In the LSND beam, the neutrinos cross about 10 m of copper and steel. This 
corresponds to a path weighed by the square electron density of the traversed 
matter of approximately 130 m ($gcm^{-3})^2$.
In a simplified description, the earth is composed of a central core of radius
3500 km and density 11.5 $gcm^{-3}$, surrounded by a mantel of 3000 km 
thickness and density 4.5 $gcm^{-3}$. This gives, for a neutrino crossing 
the whole diameter of the earth, a weighed path of about 160000 $km(gcm^{-3})^2$.
We keep the simple formula for the decay probability:

$$P = exp(-lm/Ec\tau_m)$$

where $l$ is the actual length, and $\tau_m$ includes the matter effect. 
Note that, in principle, the mass of a neutrino is affected by matter 
effects and thus can vary depending on the medium. We take here a well 
defined mass m which may or may not be the vacuum value.

Scaling from the LSND result, the probability for a 1 GeV $\nu_{\mu}$ to 
decay through the earth is 0.80. The disappearance seen by SuperKamiokande 
is about 0.50. The model seems to give an excessive deficit, but the 
enhancement in matter comes from a coherent interaction on atomic electrons, 
and is different for neutrinos and antineutrinos. Atmospheric neutrinos at 
low energy have equal populations of $\nu_{\mu}$ and $\bar{\nu_{\mu}}$. 
Because of the reduced cross-section of $\bar{\nu_{\mu}}$ , 1/4 of the 
events coming from this source are unaffected. Furthermore, the weighed 
path decreases very rapidly with the zenith angle, as the dense matter is 
concentrated in the core. For a cos$\theta$ = -0.8 (the last bin in the 
SuperKamiokande notation) the probability of decay goes down to 0.30.

With the angular resolution of SuperKamiokande, and considering the 
unaffected contribution of antineutrinos, the deficit obtained for 
contained events (sub-GeV as well as multi-GeV) is satisfactory.

The decay results in $\bar\nu_{e}$ and gives an excess of e-like events. 
However because of the reduced cross-section of antineutrinos, this excess 
is small, and can be seen in the data.

The difficulty may arise with up-going muons. Here the direction is well 
reconstructed and the model predicts a deficit of 0.07 for 5 GeV $\nu_{\mu}$ 
and 0.02 for 10 GeV $\nu_{\mu}$ between horizontal and vertical directions. 
This seems low compared with the observations.

\section{Conclusion}

The conjecture of a common origin for the LSND and SuperKamiokande findings 
is suggested. It is surprising that both experiments can be interpreted by 
the radiative decay:

 $$\nu_{\mu} \rightarrow \bar{\nu_{e}} + \gamma$$

with degenerated neutrino masses. Within this hypothesis, LSND sees the 
appearance of $\bar\nu_{e}$, while SuperKamiokande sees the disappearance 
of $\nu_{\mu}$ . Taking into account the amplification in matter, one finds 
that the lifetime inferred from LSND reproduces adequately the size of the 
effect seen in atmospheric neutrinos, at least for the contained event sample.
This lifetime is not in contradiction with other experimental results. 
A careful $\chi^2$ analysis would probably prefer the oscillation 
interpretation, but the present observation has the advantage of explaining 
both LSND and SuperKamiokande with the same phenomenon.

If the energy term in the amplification factor is the one found in Ref.[6],
the effect would be very small in the MiniBoone and 
I216 experiments proposed to check the LSND signal, and also in the high 
energy long base-line projects. On the other hand, other experimental tests 
are possible and are being studied.

\end{document}